\documentclass[reprint,amssymb,amsmath,aip,cha]{revtex4-1}
\usepackage{graphicx}
\usepackage[caption = false]{subfig}
\usepackage{color}
\usepackage{epsfig}
\usepackage{ifpdf}
\usepackage{url}%
\usepackage{cleveref}
\usepackage{bm}
\usepackage[colorlinks=true,linkcolor=blue]{hyperref}%
\expandafter\ifx\csname package@font\endcsname\relax\else
 \expandafter\expandafter
 \expandafter\usepackage
 \expandafter\expandafter
 \expandafter{\csname package@font\endcsname}%
\fi
\begin{document}
\title{Collective dynamics of large aspect ratio dusty plasma in an inhomogeneous plasma background: Formation of the co--rotating vortex series}
\author{Mangilal Choudhary}
\email{mangilal@ipr.res.in}
\affiliation{Institute for Plasma Research, Bhat, Gandhinagar, 382428, 
India}
\affiliation{Homi Bhabha National Institute, Training School Complex, Anushakti Nagar, Mumbai 400085,
India}
\author{S. Mukherjee}
\affiliation{Institute for Plasma Research, Bhat, Gandhinagar, 382428, 
India}
\author{P. Bandyopadhyay}
\affiliation{Institute for Plasma Research, Bhat, Gandhinagar, 382428, 
India}
%
\begin{abstract}
In this paper, the collective dynamics of the large aspect ratio dusty plasma is studied over a wide range of discharge parameters. An inductively coupled diffused plasma, which creates an electrostatic trap to confine the negatively charged grains, is used to form a large volume (or large aspect ratio) dusty plasma at low pressure. For introducing the dust grains into the potential well, a unique technique using a secondary DC glow discharge plasma is employed. The dust dynamics is recorded in a 2-dimension (2D) plane at a given axial location. The dust fluid exhibits wave like behavior at low pressure (p$<$0.06 mbar) and high rf power (P$>$3 W). The mixed motion, waves and vortices, are observed at an intermediate gas pressure(p$\sim$ 0.08 mbar) and low power (P$<$3 W). Above the threshold value of gas pressure (p$>$0.1 mbar), the clockwise and anti-clockwise co-rotating vortex series are observed on the edges of the dust cloud, whereas the particles in central region show the random motion. These vortices are only observed above a threshold width of the dust cloud. The streaming ions are considered the available free energy source to excite the waves in dust grain medium. The occurrence of the co-rotating vortices is understood on the basis of the charge gradient of dust particles which is orthogonal to the gravity. The charge gradient is a consequence of the plasma inhomogeneity from the central region to the outer edge of dust fluid. Since, a vortex has the characteristic size in the dissipative medium; therefore, a series of the co-rotating vortex on the both sides of dusty plasma is observed. The experimental results on the vortex formation and its multiplicity are compared to an available theoretical model and are found to be in close agreement.  
\end{abstract}
\maketitle
\section{Introduction}
The dusty plasma, which is an admixture of the electrons, ions, neutrals, and sub--micron to micron sized negatively charged solid particles, has been a current topic of research due to its applications in space plasmas \cite{goertzdustysolarsystem,geortzspokes2,cosmicdustymendis}, plasma processing technologies \cite{selwynprocessing1,watanableprocessingplasma2}, biological systems \cite{bacteriadisruption}, condensed matters \cite{thomasdustycrystal1,thomasphasetransition1} etc. In the background of plasma, the highly mobile electrons and slower ions impinge on the dust grain surface and make it negatively charged. Thses grains are either externally introduced or internally grown in the plasma. In the low-temperature plasma, dust grains get negative charges up to $10^3-10^5$ times of an electron charge (e). The collection of these highly negatively charged grains exhibits the collective dynamics similar to the conventional two component plasmas. The result of the collective response of the dusty plasma medium is encountered as dust-acoustic modes \citep{daw2,daw3,dasw,pdasw,exp1dasw,mangilalpop,dlw1,dlw2} and vortex motion \citep{vortexmicrogravity,largescalevortices,bellanicedustyrotation,mangirotationpop}. These modes are sometimes spontaneously excited in the dust grain medium when there is a free energy source to drive the grains motion. In the laboratory dusty plasma, the streaming ion \citep{instability1,instability2,instability3} and charge gradient of dust grains \citep{selfexcitedmotioninhomogeneus,vaulinajetp,mangirotationpop} are considered mainly free energy source to compensate the dissipation losses. For the steady motion of dust grains, energy dissipation losses should be minimum so that small amount of available energy can trigger the instabilities. The evaluation of such instabilities give rise to dust acoustic waves \citep{mangilalpop,dawmerlino} and vortex structures \citep{vaulinajetp,selfexcitedmotion,mangirotationpop}. It has been observed in many experiments that instabilities to excite the waves in dusty plasma is strongly dependent on the dissipation losses, i.e., dust--neutral collisions. The excitation of such low-frequency acoustic modes is only possible below the critical friction frequency \cite{daw2,daw3,pramanikddw,mangilalpop}. On the other hand, the convective or rotational motion of dust grains is sometimes observed independent of the dissipation losses of the dust grain medium \cite{vaulinajetp,selfexcitedmotioninhomogeneus,horizontalrotation,manjeetrotation}. In the absence of magnetic field, the source of convective or rotational motion of dust grains is asymmetric ion flow along with the electric field \citep{rotationinionflow,vortexmicrogravity,laishramshearflow}, or charge gradient of the dust particles along with the non--electrostatic forces \citep{vaulinajetp,vaulinaselfoscillation,selfexcitedmotion,zhdanovnonhamiltonian}, or Rayleigh--Taylor instability \citep{rtinstabilityvortices} or transient shear instability \citep{transientinstabilityrotation} or the convective motion of background neutral gas\citep{thermalconvection,thermalcreeprotation}. The studies on rotational or vortex motion of dust grains in absence of magnetic field such as spontaneous rotation of dust particles discharge \citep{Agarwalrotation}, two--dimensional (2D) dust vortex flow \cite{uchida2dflow}, cluster rotation \cite{clusterrotationunmagnetizedplasma}, horizontal and vertical vortices in presence of an auxiliary electrode \citep{horizontalrotation,probeinducedcirculation}, poloidal rotation of dust grains with toroidal symmetry\citep{manjeetrotation}, and vortex motion along with waves \cite{selfexcitedmotioninhomogeneus,selfexcitedmotion,inductivelycoupledrotation} are carried out in the various dusty plasma devices.\par
In recent studies, the co-rotating vortices are observed in an extended unmagnetized dusty plasma \cite{mangirotationpop}. The charge gradient of dust grains along with the ion drag force, which is due to the plasma inhomogeneity along the dust cloud axis, is considered the source of vortex flow \cite{mangirotationpop,mangilalrsi}. These observed exciting results on the co-rotating vortices have been created an interest to study the large aspect ratio dusty plasma in inhomogeneous plasma background.  
\par
The present study focuses on the collective dynamics of large aspect ratio dusty plasma medium, which is produced in the potential well of inductively coupled diffused plasma. The dusty plasma exhibits self-oscillatory motions such as acoustic vibration (waves) and rotational motion (vortices) at different discharge conditions. At higher rf power and low pressure, the self-excited dust acoustic waves are observed. At low power and intermediate gas pressure, the central region of dust cloud exhibits acoustic waves whereas edge particles participate in the vortex motion. At higher gas pressure, the acoustic vibrations in the central region of dust cloud diminish and particles show random motion. At this discharge condition, the particles at edges of the dust cloud rotate in the clockwise and anti-clockwise direction respectively and form a series of co-rotating vortex structures. The dependence of friction frequency on the vortex motion is studied at given input rf power. These experimental results on the vortex motion of dust grains and its multiplicity are compared to an available theoretical model \cite{vaulinajetp,selfexcitedmotioninhomogeneus} and found to be in good agreement. 
\par


The manuscript is organized as follows: Section~\ref{sec:exp_setup} deals with the detailed description of the experimental set-up, plasma, and dusty plasma production. The particles confinement in the potential well of diffused plasma is discussed in Section~\ref{sec:confinment_dust_plasma}. The detailed characteristics of the large aspect ration dusty plasma at various discharge parameters are presented in Section~\ref{sec:dynamics_dusty_plasma}. Quantitative analysis of origin of vortex flow and its multiplicity in the dusty plasma medium is described in Section~\ref{sec:explanation_vortex_formation}. A brief summary of the work along with concluding remarks is provided in Section~\ref{sec:conclusion}.
\section{Experimental Setup and Diagnostics} \label{sec:exp_setup}
\begin{figure*}
\centering
  \includegraphics[scale=0.80]{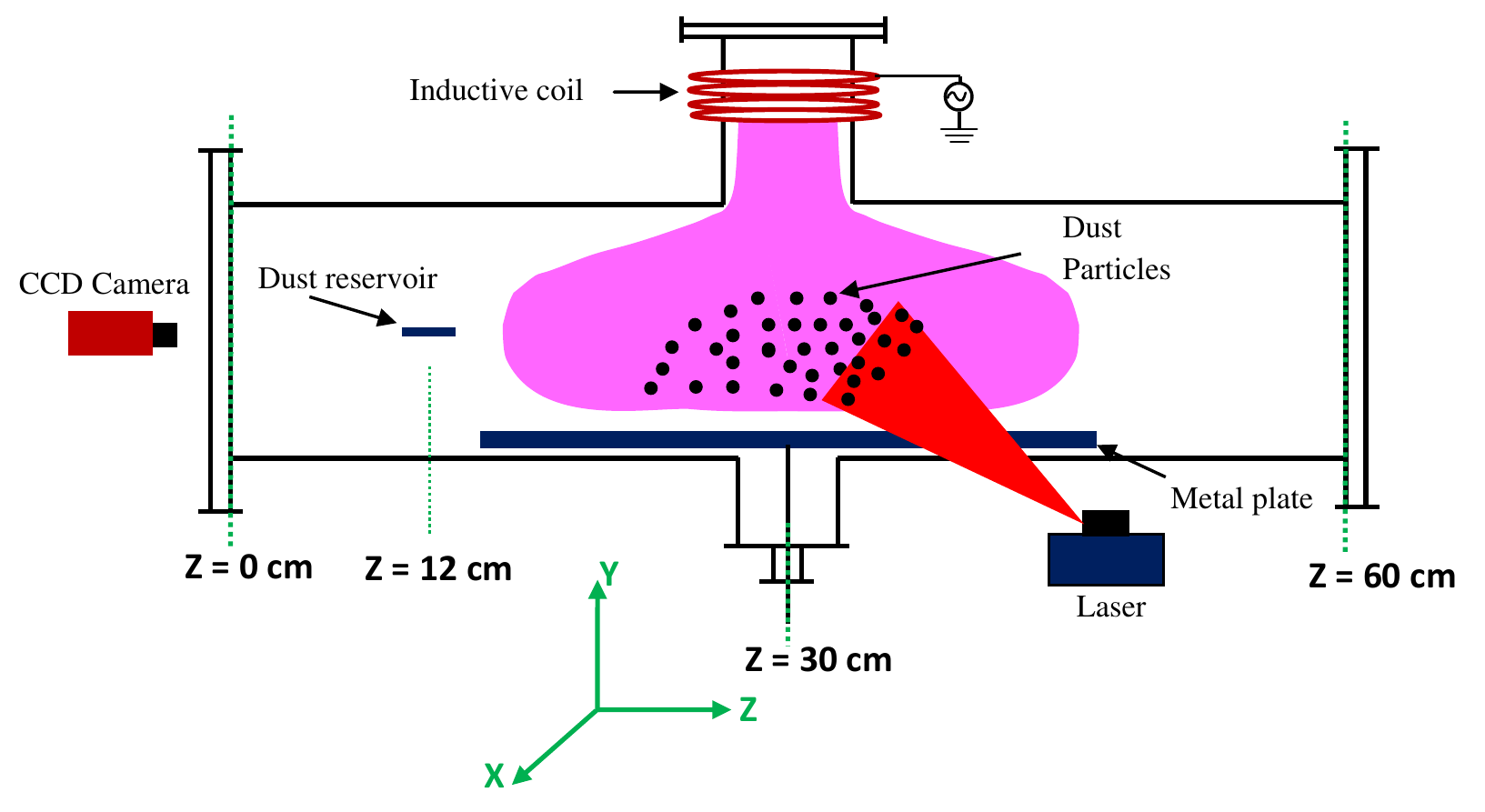}
\caption{\label{fig:fig1} Schematic diagram of the experimental setup for the study of large aspect ratio dusty plasma.}
\end{figure*}

A cylindrical linear device made of borosilicate glass tube with inner diameter of 15 $cm$ and length of 60 $cm$ is used to carry out the experimental studies on the collective phenomena in a large volume (or large aspect ratio) dusty plasma. The detailes of the experimental setup is discussed elsewhere \cite{mangilalrsi}. The schematic diagram of the experimental configuration to produce the large volume dusty plasma is shown in Fig.~\ref{fig:fig1}. 
\par
In this particular experimental configuration, Z = 0 $cm$ and Z = 60 $cm$ correspond to the left and right axial ports (see Fig.~\ref{fig:fig1}), respectively. X = 0 $cm$ and Y = 0 $cm$ indicate the points on the axis passes through the center of the experimental chamber. The center of the source tube (radial port) is located at Z $\sim$ 30 $cm$, whereas the dust reservoir (a stainless steel disk of 6 cm diameter with a step-like structure of 5 $mm$ width and 2 $mm$ height at its periphery) is mounted at one of the radial ports (at Z $\sim$ 12 $cm$) of the chamber. The dust particles are homogeneously sprinkled on the disk surface (reservoir), which is located inside the experimental chamber (at Z $\sim$ 12 $cm$). The experimental chamber is evacuated up to $\sim 10^{-3}~mbar$ by using a rotary pump. Afterwards, the argon gas is fed into the chamber till the pressure attains the values of $\sim$ 4--5 $mbar$. Then, the chamber is again evacuated to the base pressure. This process is repeated five to six times to reduce the impurities of air from the vacuum chamber. Finally, the operating pressure is set between the range of 0.05 to 0.3 $mbar$ by precisely adjusting the gas dosing valve.\par

A loop antenna (4 turns of enamel copper wire) is wound on the cylindrical source tube (8 $cm$ long and 7.5 $cm$ diameter) as indicated in Fig.~\ref{fig:fig1}. The discharge is initiated in the background of argon gas in the source tube using a 13.56 MHz rf generator. This rf plasma diffuses in the main experimental chamber. The diffused plasma in the main experimental chamber is characterized thoroughly by using different electrostatic probes namely,  single \cite{probemerlino}, double \cite{doubleprobemalter} Langmuir and emissive \cite{emissivesheehan} probes. 
\par
For injecting kaolin dust particles ($\rho_d \sim$ 2.6 $gm/cm^3$ and $r_d \sim$ 0.5 to 4 $\mu$m) into the electrostatic trap, the dust reservoir is biased negatively ($\sim$ -300 V or above) to form a secondary DC glow discharge around the disk. In the background of this secondary plasma, the dust particles get negatively charged and lifted up near the plasma--sheath boundary. Since, the particles are poorly confined in the cathode sheath region at low pressure; therefore, they continuously leave the dust reservoir. As these particles come into diffused plasma, they start to flow towards the center of source tube and are found to confine in the potential well of diffused plasma near the source section (at Z $\sim$ 30 cm). An ambipolar E--field of the diffused plasma is reposnsible to transport and trap the particles \cite{mangilalrsi}.
 \par
The confined particles are then illuminated in the X--Y plane by the combination of a tunable red diode laser (632 nm wavelengths, 1--100 $mW$ power and $\sim$ 3 $mm$ beam diameter) and a cylindrical lens, whereas the dynamics of the dust grains are captured by a CCD camera. The stored images are analyzed with the help of ImageJ \cite{imagejsoftware} software and MATLAB based openPIV \cite{piv} software. \par 
\section{Levitation and confinement of the dust particles} \label{sec:confinment_dust_plasma} 
In the background of diffused plasma, micronsized dust grains get negatively charged by collecting more elelctrons than ions. These negatively charged dust grains get confined under the action of electrostatic force, gravitational force, ion drag force and neutral drag force. In the present experimental configuration, neutrals only resist the motion of dust grains; therefore, its role in confinement is not considered. The role of ion-drag force on the dust grains is found to be less dominated than an electrostatic force for the given experimental parametric regime \cite{mangirotationpop}. It essentially means that gravitational and electrostatic force dominates over the other forces. For understanding the confinement, plasma potential is measured along the X, Y and Z directions.
 \begin{figure*}
 \centering
\subfloat{{\includegraphics[scale=0.315]{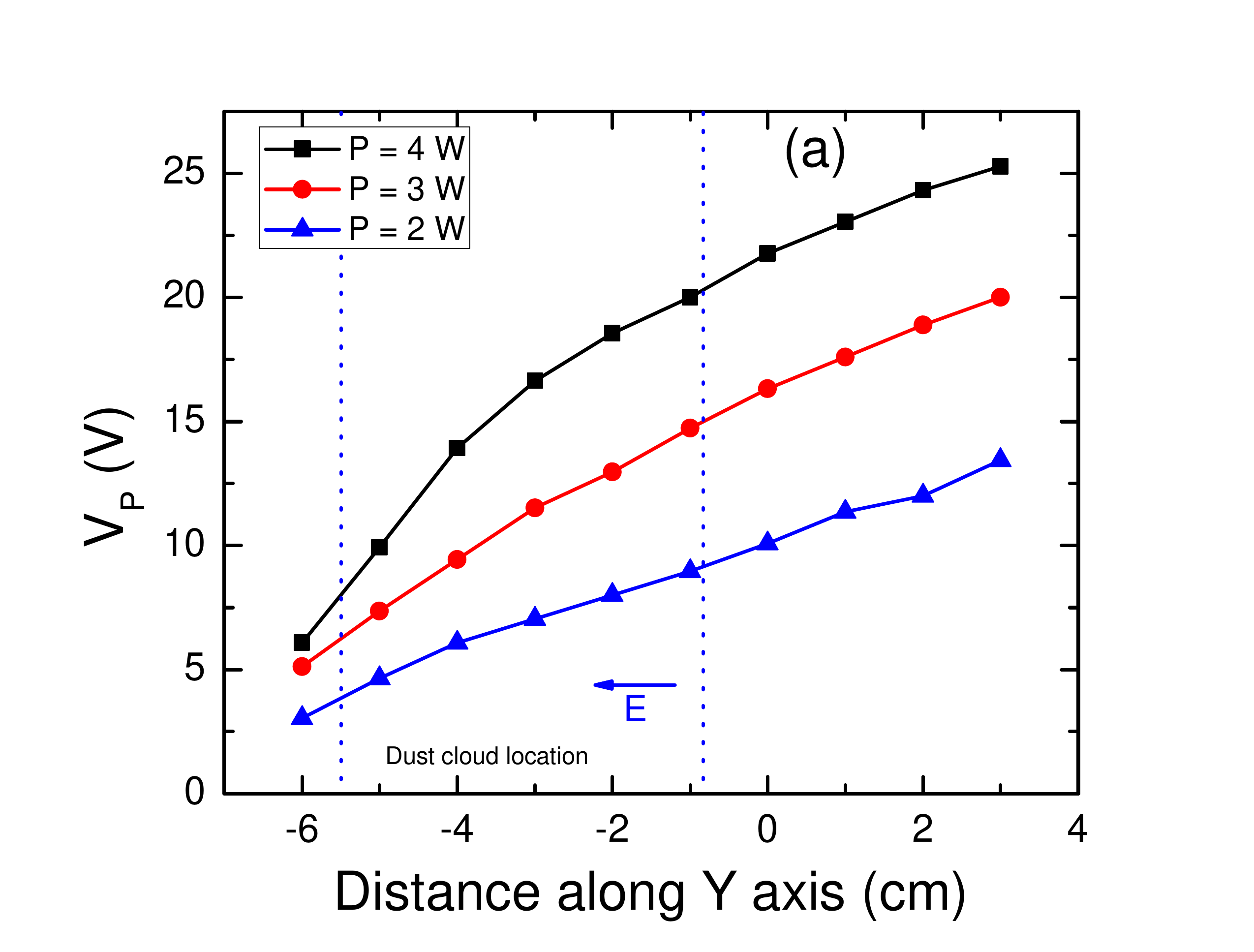}}}%
 \qquad
 \subfloat{{\includegraphics[scale=0.3150]{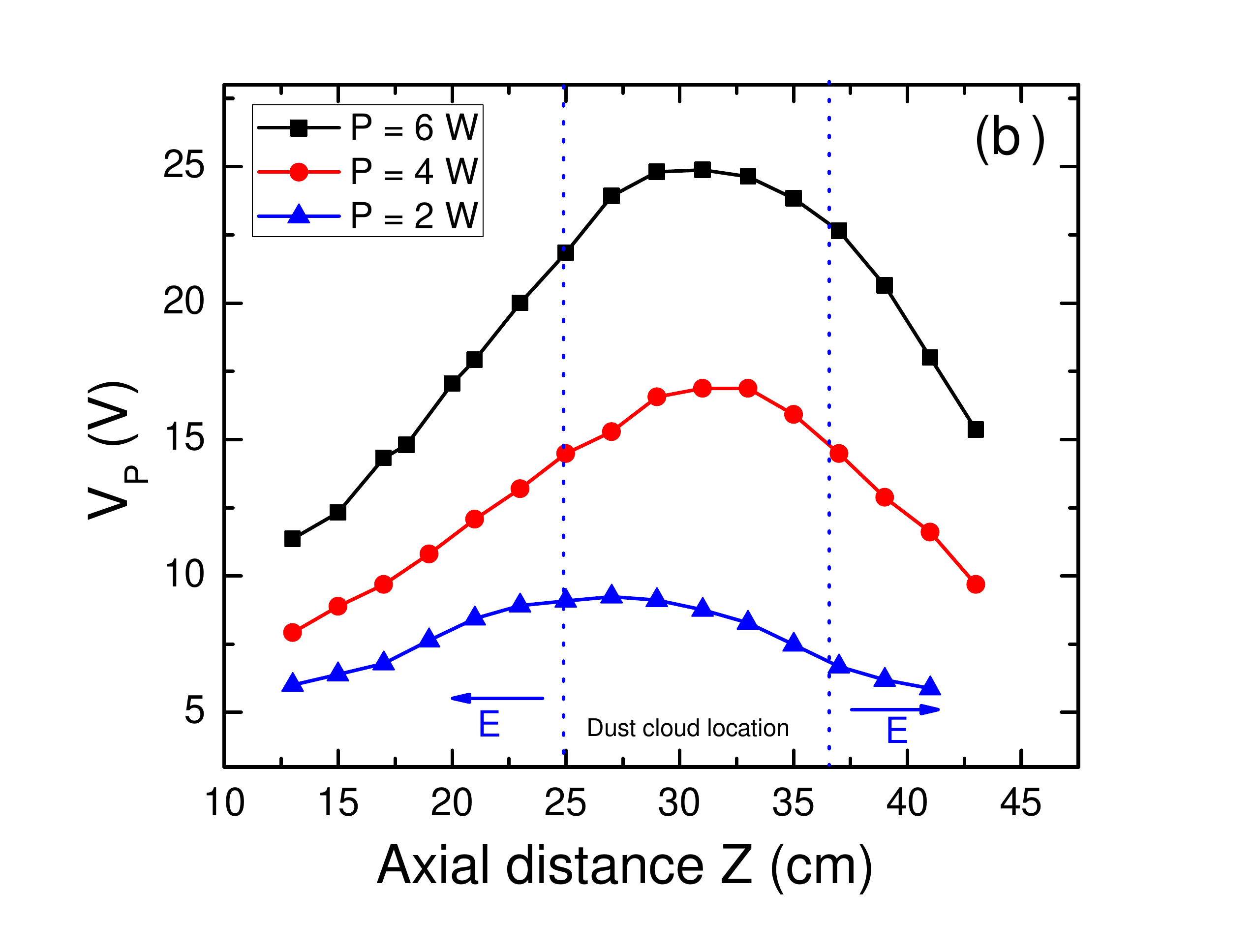}}}
 \qquad
\subfloat{{\includegraphics[scale=0.315]{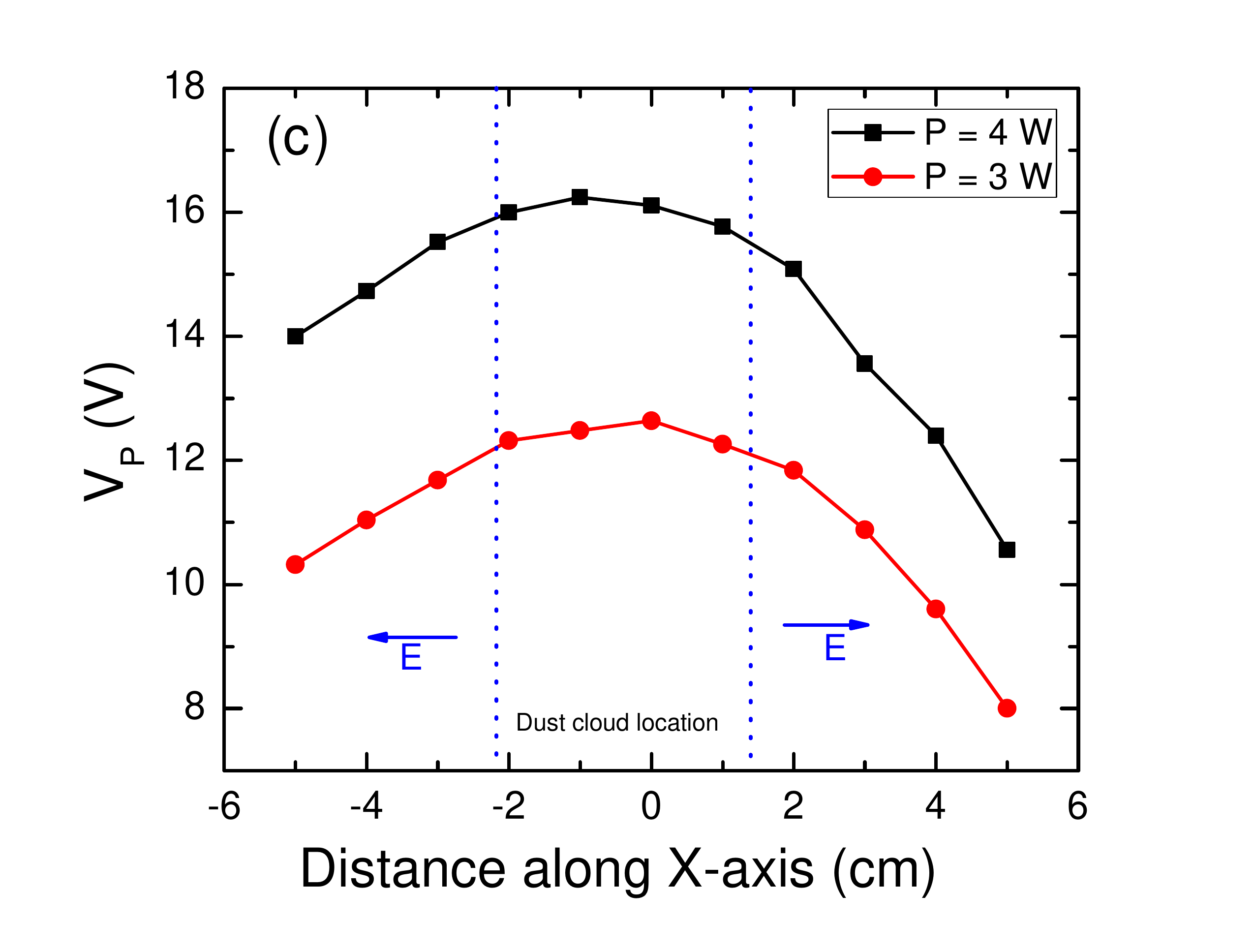}}}%
 \caption{\label{fig:fig2}(a) Plasma potential ($V_P$) variation along the Y--axis at X $\sim$ 0 $cm$ and Z $\sim$ 30 $cm$ for rf powers, P = 4 W, 3 W and 2 W. (b) $V_P$ variation along the Z--axis at Y $\sim$ -3 $cm$ and Z $\sim$ 30 $cm$ for rf powers, P = 6 W, 4 W and 2 W. (c) $V_P$  variation along the X--axis at Y $\sim$ -3 $cm$ and Z $\sim$ 30 $cm$ for two rf powers, P = 4 W and 3 W. All the measurements are taken in the absence of the dust particles at gas pressure, p = 0.09 $mbar$. The errors in the measured value of plasma potential are within $\pm$ 5\%.} 
 \end{figure*}
\par
The levitation of massive dust grains against the gravity is understood by measuring the plasma potential along the Y-direction. The plasma potential is measured in the vertical direction (along Y-axis) at Z $\sim$ 30 cm and X $\sim$ 0 cm for different rf powers. The variation of plasma potentials in the vertical direction is depicted in Fig.~\ref{fig:fig2}(a). It is clearly seen in Fig.~\ref{fig:fig2}(a) that plasma potential has a gradient in the vertical direction, which gives rise to an E-field to hold the charged dust grains against the gravity. The levitated dust grains form an elongated dust cloud in the vertical direction. The length of the confined dust cloud in Y-direction depends on the input rf powers.
\par
Fig.~\ref{fig:fig2}(b) shows the axial (along Z-axis) plasma potential profiles for different rf powers at an argon pressure, p = 0.09 mbar. The plasma potential is measured at X $\sim$ 0 cm and Y $\sim$ -3 cm. The plasma potential is found to be higher near the center of source tube (at Z $\sim$ 30 cm) and decreases towards the edges of diffused plasma. The gradient in plasma potential gives rise to an E-field component, which is indicated by an arrow in Fig.~\ref{fig:fig2}(b). Thus, an electrostatic force due to this E-field provides the axial confinement to dust grains. An axial dimension of dust cloud is determined by the plateau region of plasma potential, which strongly depends on the input rf power (see Fig.~\ref{fig:fig2}(b)).
\par
Fig.~\ref{fig:fig2}(c) shows the plasma potential variation along the X-axis at  Z $\sim$ 30 cm and Y $\sim$ -3 cm for different rf powers. The gradient in plasma potential on the both sides gives rise to the E-field, which confines the negatively charged dust grains in the X-direction. It is also observed that dust cloud length along the X-axis (or width) depends on input rf power. The dust cloud width decreases with decreasing the input rf power, which can be determined by the plateau region of plasma potential. Hence, the ambipolar E-field of diffused plasma is responsible to levitate and confine the negatively charged dust grains.
\section{Dynamics of large aspect ratio dusty plasma} \label{sec:dynamics_dusty_plasma} 
The confined dust grains in a potential well of inductively coupled diffused plasma exhibits various types of collective phenomena such as waves, convective motion, vortex motion etc. Since, the collective dynamics of the dust grain medium is associated with the ambient plasma environment; therefore, its dynamics get modified with the change of plasma parameters. In the present experimental configuration, the dusty plasma has 3-dimensional (3D) nature however the imaging diagnostics restrict ourselves to track the dynamics in a 2D plane; therefore, dust grains dynamics is studied in an extended (or large aspect ratio) dusty plasma medium in the X-Y plane, where dusty plasma is found to be homogeneous in the axial direction. The characteristics of the dusty plasma in a vertical (in the X-Y ) plane at various discharge conditions are discussed in the following subsections. 
 \subsection{Transition from waves to vortex}
\begin{figure*}
\centering
  \includegraphics[scale= 0.9000]{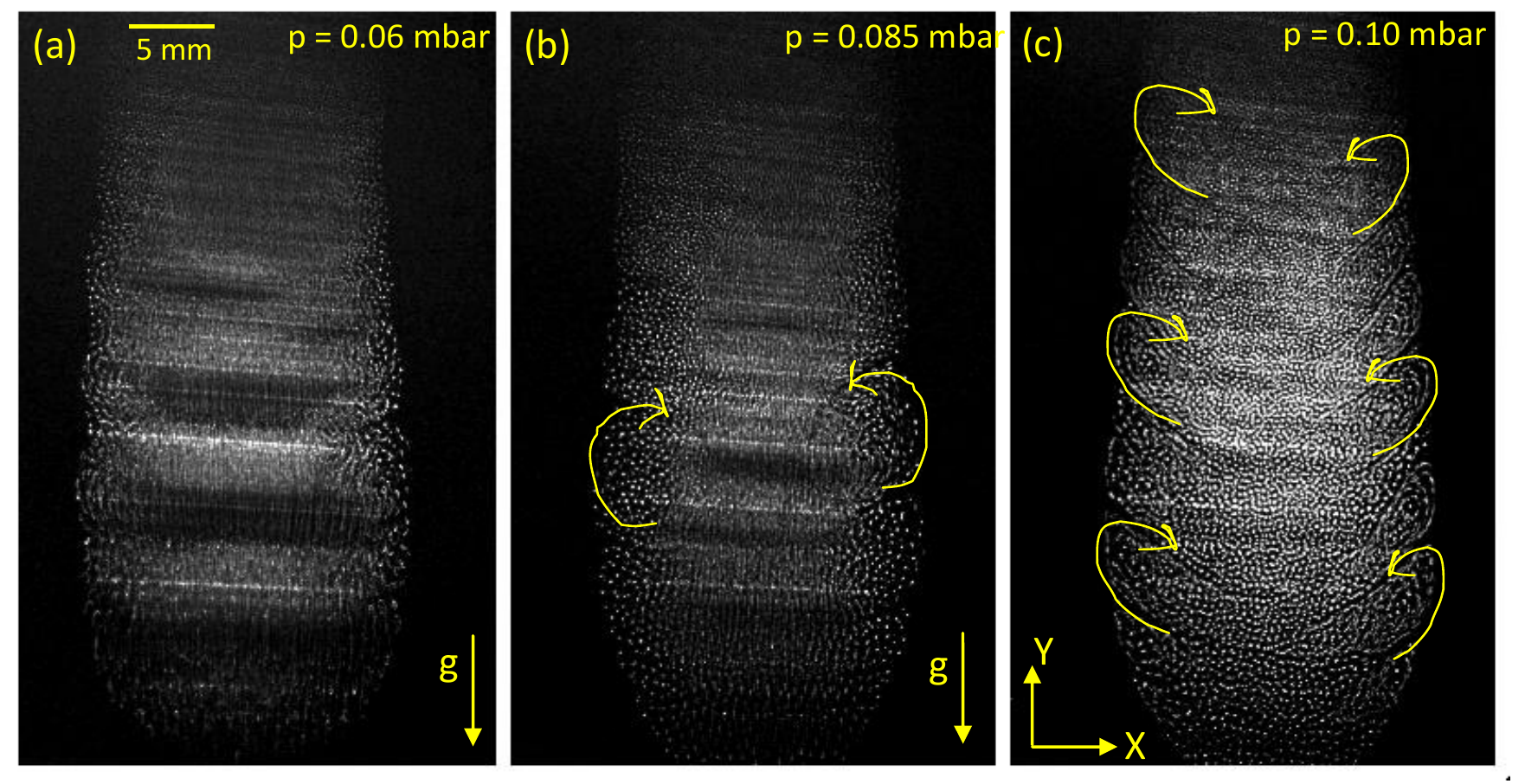}
\caption{\label{fig:fig3}Video images of the dust cloud in the X--Y plane at Z $\sim$ 30 cm. Fig.\ref{fig:fig3}(a)--Fig.\ref{fig:fig3}(c) show the dust dynamics at different neutral pressures at a fixed rf power, P = 2.8 W. The yellow solid lines with an arrow indicate the direction of vortex motion of dust grains in this plane.}  
 \end{figure*}

The dynamics of dust grains is recorded in the X-Y plane at Z $\sim$ 30 cm for different neutral pressures at a fixed input rf power, P = 2.8 W. The characteristics of an elongated dusty plasma at different gas pressure is depicted in Fig.~\ref{fig:fig3}. The dust cloud exhibits wave like motion at lower pressure, p = 0.06 mbar, (see Fig.~\ref{fig:fig3}(a)). With increasing the pressure to 0.085 mbar, the co-existence of wave and vortex motions is observed. In this conditions, the waves in the central region of dust cloud propagates along the direction of gravity, whereas at the edge of the cloud particles rotate in clockwise and anti-clockwise directions (see Fig.~\ref{fig:fig3}(b)). 

Further increase in the gas pressure to 0.10 mbar, the grains of the central region of the dust cloud have the random motion instead of the acoustic vibrations, whereas the particles at the edge of dust grain medium participate in the vortex motion and form a series of the co-rotating vortex on the both sides of the dust cloud. Since, the dust cloud is symmetric about its central axis; therefore, clockwise and anticlockwise co-rotating vortex structures are observed at this discharge condition (see Fig.~\ref{fig:fig3}(c)). The acoustic waves at less dissipation losses are attributed to ion streaming instabilities in the dusty plasma \cite{instability1, instability2,instability5,mangilalpop}.   
This transition from the waves to random motion with increasing the neutral pressure or friction frequency is due to the suppression of associated instabilities. The appearance of the vortex motion on the both sides (or edges) of dust cloud is a result of the another instability, which is discussed in the subsequent section. 

 
\subsection{Vortex dynamics with friction frequency}
\begin{figure*}
\centering
  \includegraphics[scale= 0.9500]{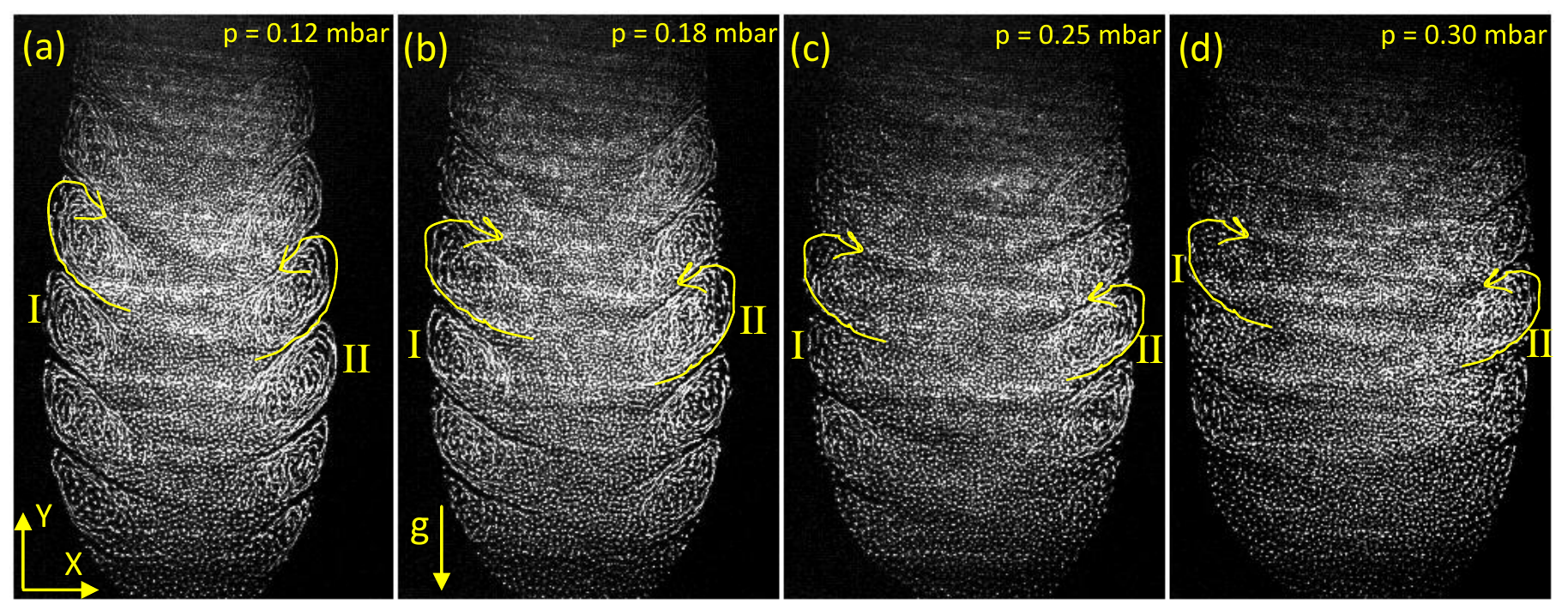}
\caption{\label{fig:fig4} Video images of the dust cloud in the X--Y plane at Z $\sim$ 30 cm. Images ((a)--(d)) are obtained by the superposition of eight consecutive images at a time interval of 66 ms. Fig.\ref{fig:fig4}(a)--Fig.\ref{fig:fig4}(d) show the observed vortex structures for different gas pressures at fixed input rf power, P = 2.8 W. The yellow solid lines with an arrow indicate the direction of vortex motion of dust grains in this plane. The vortex representation (I and II) are made for the quantitative analysis of the vortex motion.} 
 \end{figure*}

The role of the friction frequency on the vortex motion is studied at an rf power of 2.8 W. Fig.~\ref{fig:fig4} shows the dust dynamics at different neutral pressures in the X-Y plane at Z $\sim$ 30 cm. These images are constructed by the superposition of eight consecutive still images. The directed motion of dust grains forms a chain like structure, whereas the random motion of grains leaves white dots. Therefore, the co-rotating vortex structures at the edges of the cloud are results of the rotational motion of dust grains. It is to be noted that dust cloud width increases with the increase of the gas pressure due to the increase of width of the potential well as shown in Fig.~\ref{fig:fig8}(b). It is observed that edge particles have vortex motion even at higher gas pressure (p = 0.3 mbar), as is seen in Fig.~\ref{fig:fig4}(d). The shape and size of vortex depends on the dimension of dust cloud as well as plasma parameters. It is clearly seen in Fig.~\ref{fig:fig4} that two co-rotating vortices in the dust cloud medium are well separated at the interface of opposite flowing medium. With increasing the length of the dust cloud in this plane, the number of vortex structures increases. It clearly demonstrates the role of the dust cloud dimension in the formation of the series of co-rotating vortex structures. 

 \begin{figure*}
\centering
  \includegraphics[scale = 0.85]{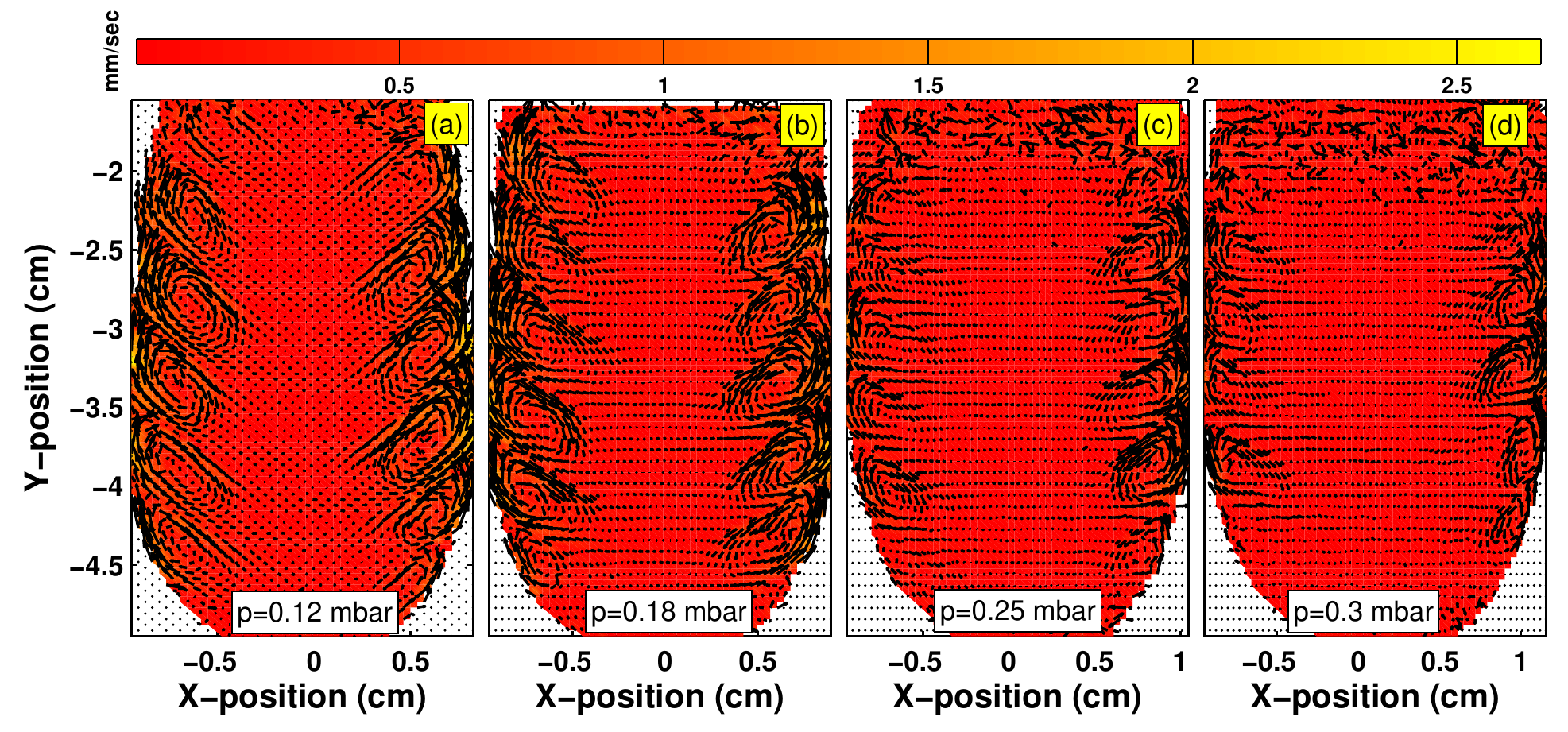}
\caption{\label{fig:fig5} Images show the velocity distribution of dust particles in a vortex structure at different gas pressures (for Fig.\ref{fig:fig3}). Images (Fig.\ref{fig:fig5}(a)--Fig.\ref{fig:fig5}(d)) are obtained after PIV analysis of the corresponding still images. Velocity vectors show the direction of the rotating particles in the X--Y plane at Z $\sim$ 30 cm. The color bar on the images show the value of dust velocity in $mm/sec$. The clockwise and anti--clockwise co--rotating vortex series are observed on the either side of dust cloud. All the measurements are taken at fixed rf power, P = 2.8 W.}  
 \end{figure*}

\paragraph*{•}
To get more information on the velocity distribution and angular frequency of the rotating grains in a vortex structure, the still images are analyzed using the MATLAB based software openPIV \cite{piv}. Fig.~\ref{fig:fig5} represents the PIV images of the dust grain medium at different neutral pressure. For constructing the vector field, an adaptive 2-pass algorithm (a 64$\times$ 64, 50\% overlap followed by a 32$\times$ 32, 50\% overlap analysis) is adopted. The contour maps of the average magnitude of the velocities are constructed after averaging the velocity vectors of 50 frames, as is shown in Fig.~\ref{fig:fig5}. The direction of the velocity vector represents the direction of rotating particles in the dusty plasma. The edge particles rotate in the clockwise on the left side and the anti-clockwise on the right side of the dust cloud in the X-Y plane, which is clearly indicated by the velocity vectors. It is clear from Fig.~\ref{fig:fig5} that velocity distribution of rotating particles is non-uniform in the vortex structures for all the parametric regimes. The particles of outer edge have higher rotational speed than the inner edge. The shape of the vortex structures are almost symmetric about an axis passes through the center of it. Also, the observed vortices are not to be in circular shape in the X-Y plane but have the elliptical or distorted elliptical shape in the X-Y plane. The trajectories of dust grains depend on the charge gradient in the X-direction as well as in the Y-direction \cite{vaulinaselfoscillation}. In present configuration, contribution of Y-component is considered to be  negligible than X-component; therefore, only X-component of charge gradient ($\beta_x = \beta$) is considered to understand the observed results. To estimate an average angular frequency of particles in the vortices at different gas pressure, the circular region of a vortex is considered. For a given discharge condition, all the co-rotating vortices on the both side of dust medium are observed to be nearly similar in size and having an almost similar velocity distribution (see Fig.~\ref{fig:fig5}). The average rotation speed of the particles decreases with the increasing of gas pressure or friction frequency, which is clearly seen in Fig.~\ref{fig:fig5}. 

\subsection{Dynamics of different widths dusty plasma}
\begin{figure*}
\centering
  \includegraphics[scale= 0.8500]{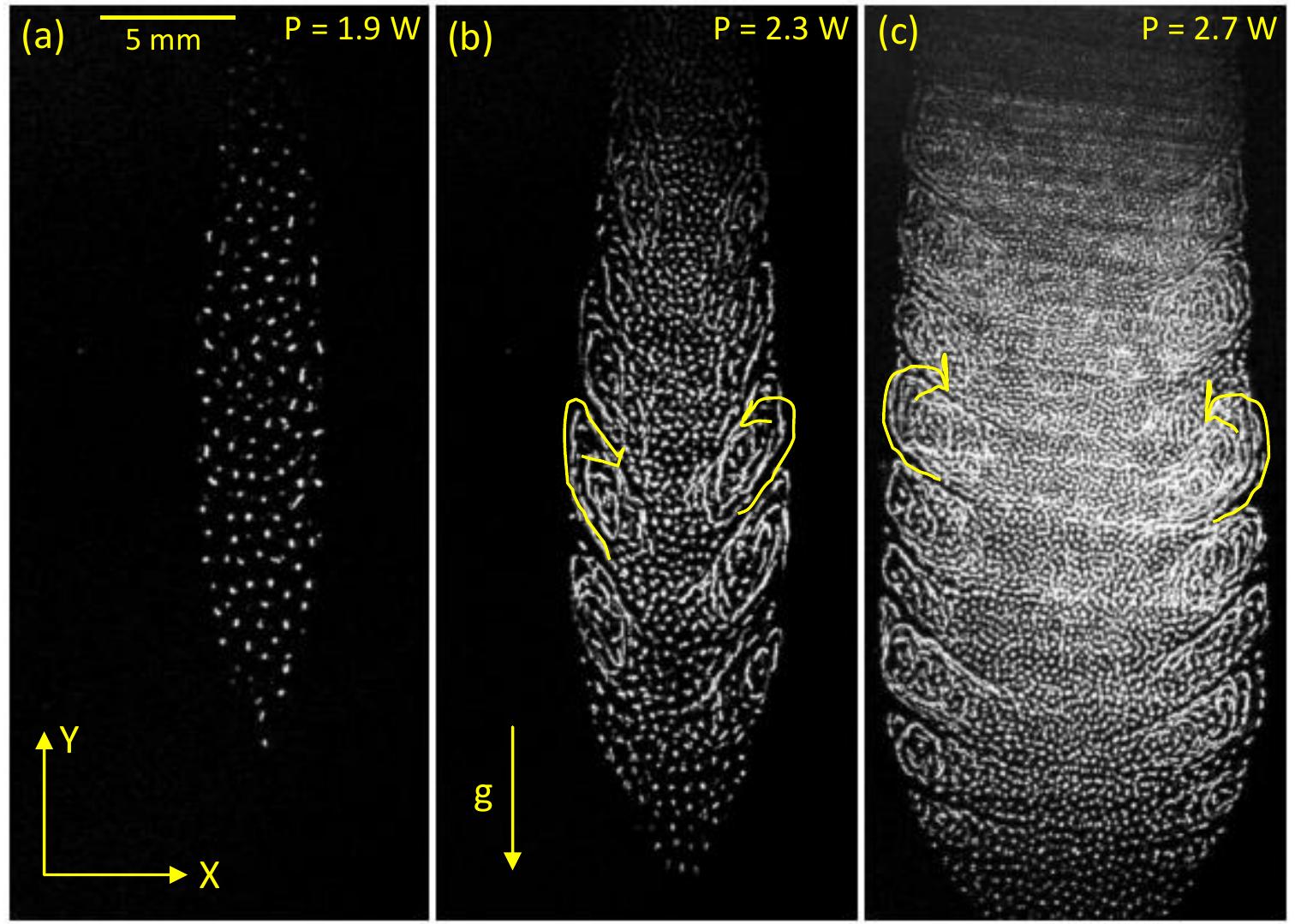}
\caption{\label{fig:fig6} Video images of the dust grain medium of different widths in the X-Y plane at Z $\sim$ 30 cm.  Images ((a)--(c)) are obtained by the superposition of eight consecutive images at a time interval of 66 ms. Fig.\ref{fig:fig6}(a)--Fig.\ref{fig:fig6}(c) show the dynamics of grains in the different widths dusty plasma. The experiments are performed at constant pressure, p = 0.10 mbar and different rf powers, P = 1.9 W, 2.3 W and 2.7 W.} 
 \end{figure*}

 Fig.~\ref{fig:fig6} shows the characteristics of the dusty plasmas of different widths (or aspect ratio) at given discharge parameters. The images in Fig.~\ref{fig:fig6} are constructed from the superposition of seven consecutive still images. The dust grains only exhibit random motion below a threshold value of the dust cloud width (or aspect ratio), as is seen in Fig.~\ref{fig:fig6}(a). Above the threshold width, the grains located at edges rotate in the clockwise and anti-clockwise directions and form a co-rotating vortex series on each side of the dust cloud. The size of the vortex is found to be dependent on the width of the dust cloud at fixed discharge conditions.

\section{Discussion of Experimental Results} \label{sec:explanation_vortex_formation} 
The existence of dust acoustics waves at higher power and lower pressure (p $<$ 0.1 mbar) is due to the instabilities, which are associated with the streaming ions in the ambipolar E-filed of the diffused plasma \cite{icpddw,instability1,instability2,mangilalpop}. Such instabilities mainly observed above a threshold E-field so that $v_i \geq v_{Ti}$, where $v_i$ and $v_{Ti}$ are the ion streaming velocity and ion thermal speed, respectively. The ions streaming velocity increases with the increase of an E-field at given pressure. The plasma potential variation at p = 0.06 mbar along the gravity is presented in Fig.~\ref{fig:fig2}(a). The estimated E-field is observed to be higher at higher power. It's value decreases with lowering the input power at given pressure. Therefore, the waves are observed at higher power and low pressure (see Fig.~\ref{fig:fig3}(a)). At higher pressure, the ion--neutral as well as dust--neutral collision frequency increases, which causes the damping of dust acoustic waves; therefore, waves motion transforms to random motion in the central region of dust cloud (see Fig.~\ref{fig:fig3}(c)). 

\par
It is obvious that a steady-state equilibrium dust vortex is formed when energy dissipations of the particles due to frequent dust--neutral collision and/or dust--dust interaction are balanced by the available free energy \cite{selfexcitedmotioninhomogeneus,mangilalpop}. In the diffused plasma, the spatial dependence of dust charge is one of the possible mechanisms to drive the vortex flow in dust grain medium \cite{vaulinajetp,vaulinaselfoscillation, selfexcitedmotion}. The monotonic variation (gradient) of particles charge in the dusty plasma occurs due to inhomogeneity in the background plasma parameters such as electrons (ions) density ($n_{e(i)}$) and/or electrons (ions) temperature ($T_{e(i)}$). It is recently experimentally varified by Choudhary \textit{et al.}\cite{mangirotationpop} in an extended dusty plasma medium with inhomogeneous plasma background. Theoretical analysis and numerical simulations show such type of vortex structures in the presence of dust charge gradient, $\vec{\beta} = \nabla Q_d = e\nabla Z_d$, orthogonal to a nonelectrostatic force $\vec{F}_{non}$ such as gravitational force ($\vec{F}_g$), or ion drag force ($\vec{F}_I$) acting on the dust particles in the dust cloud \cite{vaulinajetp,vaulinaselfoscillation,selfexcitedmotion}. In the equilibrium steady state, dust grains are stable under the balance of electric force and non-electrostatics force i.e. $Q_d E_y = M_d g$. In the presence of charge gradient ($\beta$), the curl of total force acting on the individual particle becomes non-zero due to a finite value of $\vec{\beta} \times \vec{E_y}$. In this case, the electric field ($E_y$) does the positive work in compensating the dissipative energy losses only when charge gradient is non zero and is orthogonal to force $F_{non}$. The combined action of E-field force and non-electrostatic force pumps the energy to vortex motion against the dissipation losses. The role of non--electrostatic forces ($\vec{F}_{non}$) in the formation of vortex structure in the dusty plasma is determined by their capacity to hold the dust grains in the region of the non-zero electric field.
\par
In recent years, Vaulina \textit{et al.} \cite{vaulinajetp, vaulinaselfoscillation,selfexcitedmotion} have performed extensive theoretical and numerical works to explain the self-oscillatory motion of dusty plasma with inhomogeneous plasma background. They predicted two types of instabilities in the dust grain medium in presence of charge gradient results from the plasma inhomogeneity. The first type of instability is named as dispersive instability and other is termed as dissipative instability. The evaluation of dispersive instability gives rise to acoustic waves, which strongly depends on the friction frequency. On the other hand, an evaluation of the dissipative instability gives rise to regular dynamic structures (vortices) and it is independent of the friction frequency ($\nu_{dn}$). Although, the angular frequency ($\omega$) of dust grains decreases with increasing the friction frequency but the qualitative nature of the dust grains medium remains unchanged. It essentially means that vortex motion of the grains in presence of charge gradient is independent of the dissipation losses of the medium.
\par
\begin{figure}
\centering
  \includegraphics[scale= 0.800]{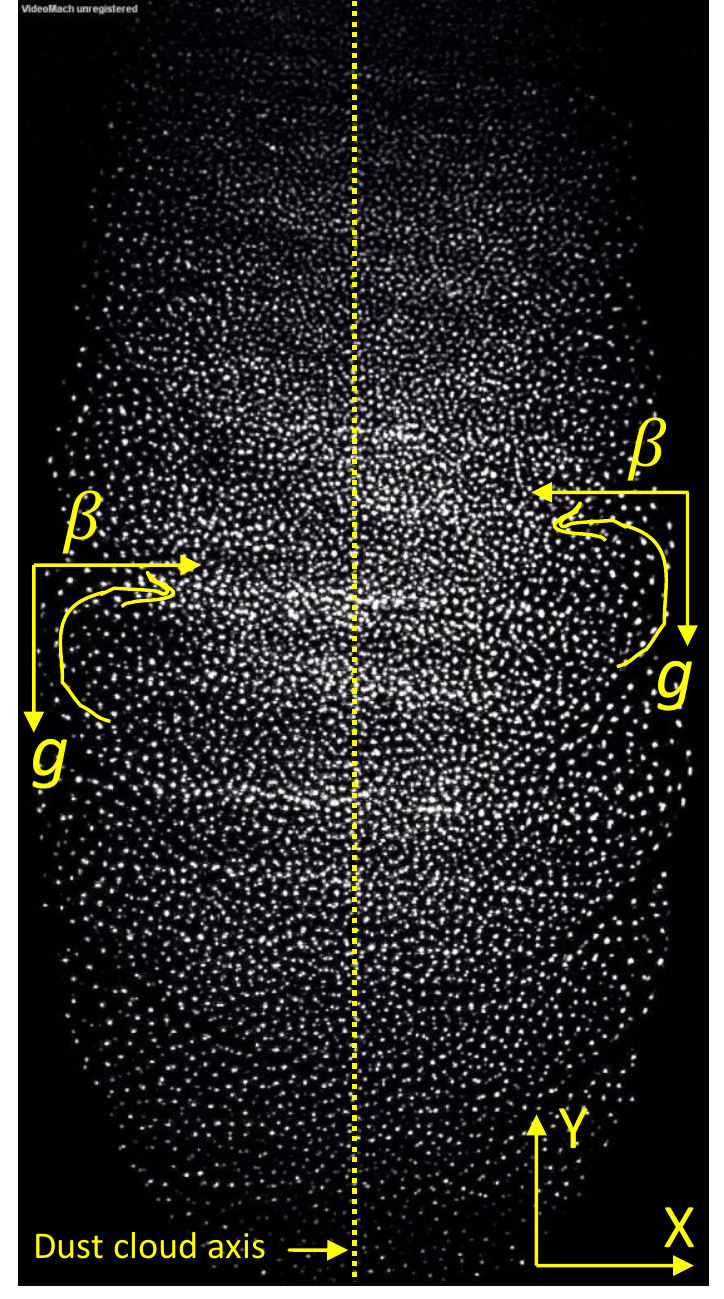}
\caption{\label{fig:fig7} Video image of dust cloud in the X--Y plane with the direction of charge gradient ($\beta$) and gravity ($g$). The direction of rotation is represented by a yellow line with an arrow. The dust grains rotate in the direction of charge gradient on either side of dust cloud.}  
 \end{figure}

Due to the dissipative instability \cite{vaulinajetp,selfexcitedmotioninhomogeneus}, dust particles in the cloud start to move in the direction of $F_{non}$ where the particle has its maximum charge value and form a vortex structure. In the vortex motion, the vorticity ($\Omega = \nabla \times \vec{v}_d$) is always non-zero along a certain closed curve. The frequency ($\omega$) of the steady-state rotation of particles in a vortex structure is given as \cite{vaulinajetp,vaulinaselfoscillation,selfexcitedmotion},
 \begin{equation}
 \omega_{th} = \vert\frac{F_{non}}{M_d} \frac{\beta}{2 e Z_0 \nu_{dn}} \vert ,
 \end{equation}
 where $e Z_0 = Q_{d0} $ is the charge on the dust particle at an equilibrium position in the rotating plane (in the X-Y plane). In our experimental configuration, the dust grains are confined in a X--Y plane at a given axial location; therefore, the  non--electrostatic force $\vec{F}_{non}$ required for the vortex motions of grains is provided by the gravitational force, i.e., $\vec{F}_{non} = \vec{F}_{g}$ \cite{selfexcitedmotioninhomogeneus,vaulinajetp}. Hence $\vec{F}_{non}$ can be replaced by $\vec{F}_g$ in the Eq.(1) to obtain the angular frequency of the rotation. So the angular frequency can be written as  
 \begin{equation}
 \omega_{th} = \vert\frac{g \beta}{2 e Z_0 \nu_{dn}} \vert ,
 \end{equation} 
It should be noted that the force experienced by the particle due to the ion drag is also orthogonal to charge gradient but the gravitation force is dominated over it; therefore, its role in the vortex motion is not considered in the calculations. Schematic diagram to represent the direction of rotational motion in the presence of charge gradient ($\beta$) and gravity in the X--Y plane is depicted in Fig.~\ref{fig:fig7}.

 The charge on the dust grain ($Q_d$) is calculated using the Matsoukas and Russel's approximations \citep{dustchargerusselapproxi.}, which is given as
\begin{equation}
\centering
Q_d = e Z_d \approx C \frac{4 \pi r_d k_B T_e}{e^2} ln \frac{n_i}{n_e}\left(\frac{m_e T_e}{m_i T_i}\right)^{\frac{1}{2}} ,
\end{equation}
 where $r_d$ is radius of the micro-particle, $k_B$ is Boltzmann's constant, $e$ is the electron charge, $n_e$ and $n_i$ are the electron and ion densities, $m_e$ and $m_i$ are their masses, and $T_e$ and $T_i$ are their temperatures. For a typical argon plasma, the constant $C$ comes out to be $\approx$ 0.73 \citep{dustchargerusselapproxi.} \par 
In the present experimental configuration, the directed gas flow inside the chamber is negligible \cite{mangilalrsi} thus neutrals are assumed to be in thermal equilibrium. According to Epstein friction \cite{neutraldragepstein}, the neutral friction experienced by the dust grains is
 \begin{equation}
 \centering
 \vec{F}_n = -m_d \nu_{dn} \vec{v}_d,
 \end{equation}
 where $\nu_{dn}$ is the dust--neutral friction frequency and $v_d$ is dust particle velocity. The expression for $\nu_{dn}$ \cite{shukladustybook} is
 \begin{equation}
 \centering
 \nu_{dn} = \frac{8}{3} \sqrt{2 \pi} r^2_d \frac{m_n}{m_d} n_n v_{Tn} \left(1 +\frac{\pi}{8} \right) ,
\end{equation}  
 where $m_n$, $n_n$, and $v_{Tn}$  are the mass, number density, and thermal velocity of the neutral gas atoms, respectively.
\paragraph*{•}
To estimate the angular frequency ($\omega$) and its dependence on the friction frequency ($\nu_{dn}$), the charge gradient ($\beta$) orthogonal to the dust cloud axis (or along the X-axis) is estimated for the given discharge conditions. The dust cloud axis is assumed along the Y-direction, as is represented in Fig.~\ref{fig:fig7}. The plasma parameters such as $n_e$, $T_e$ and $V_p$ are experimentally measured to estimate the charge gradient of dust grains along the X-direction.
  \begin{figure*}
 \centering
\subfloat{{\includegraphics[scale=0.335]{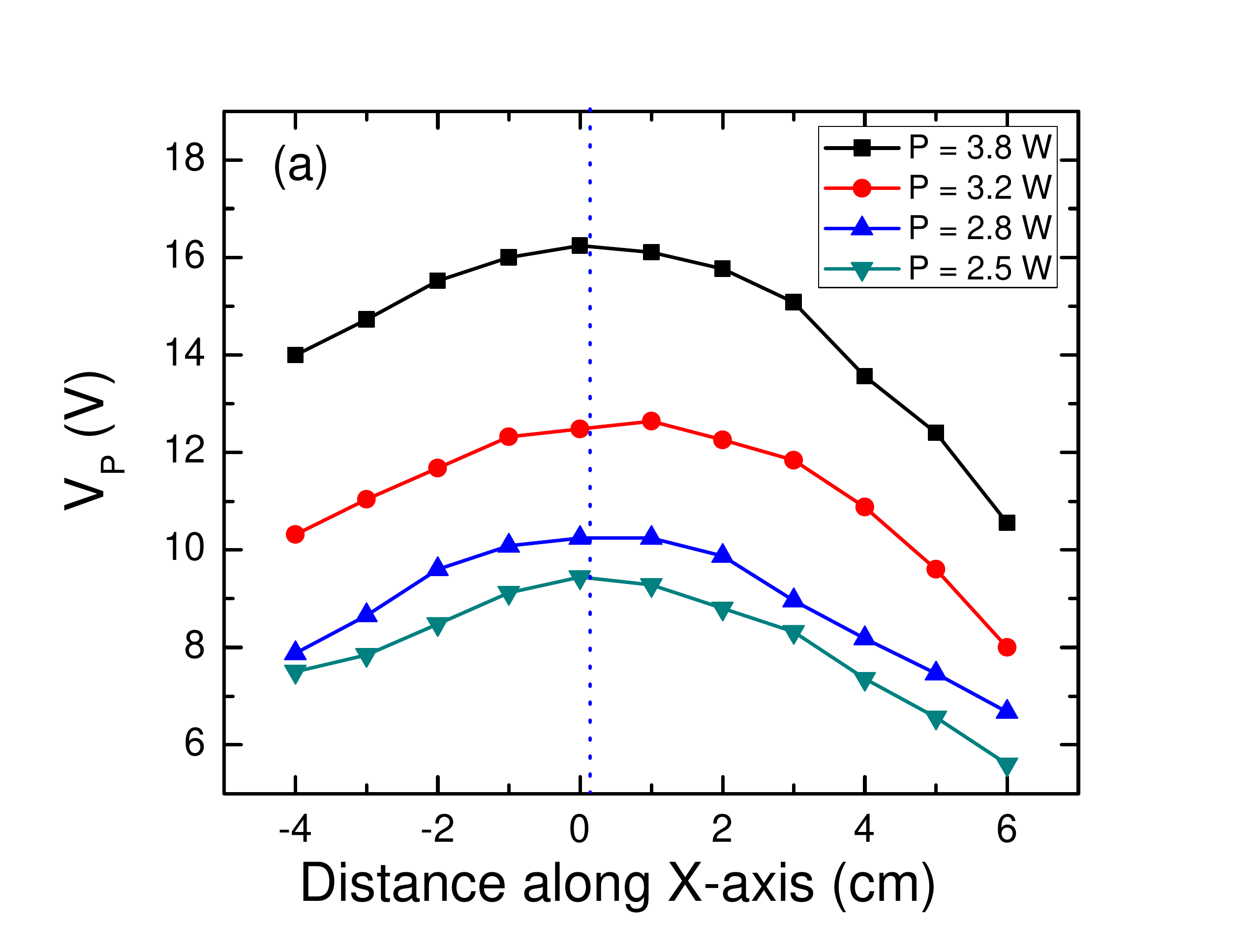}}}%
\hspace*{-0.7in}
 \qquad
\subfloat{{\includegraphics[scale=0.32]{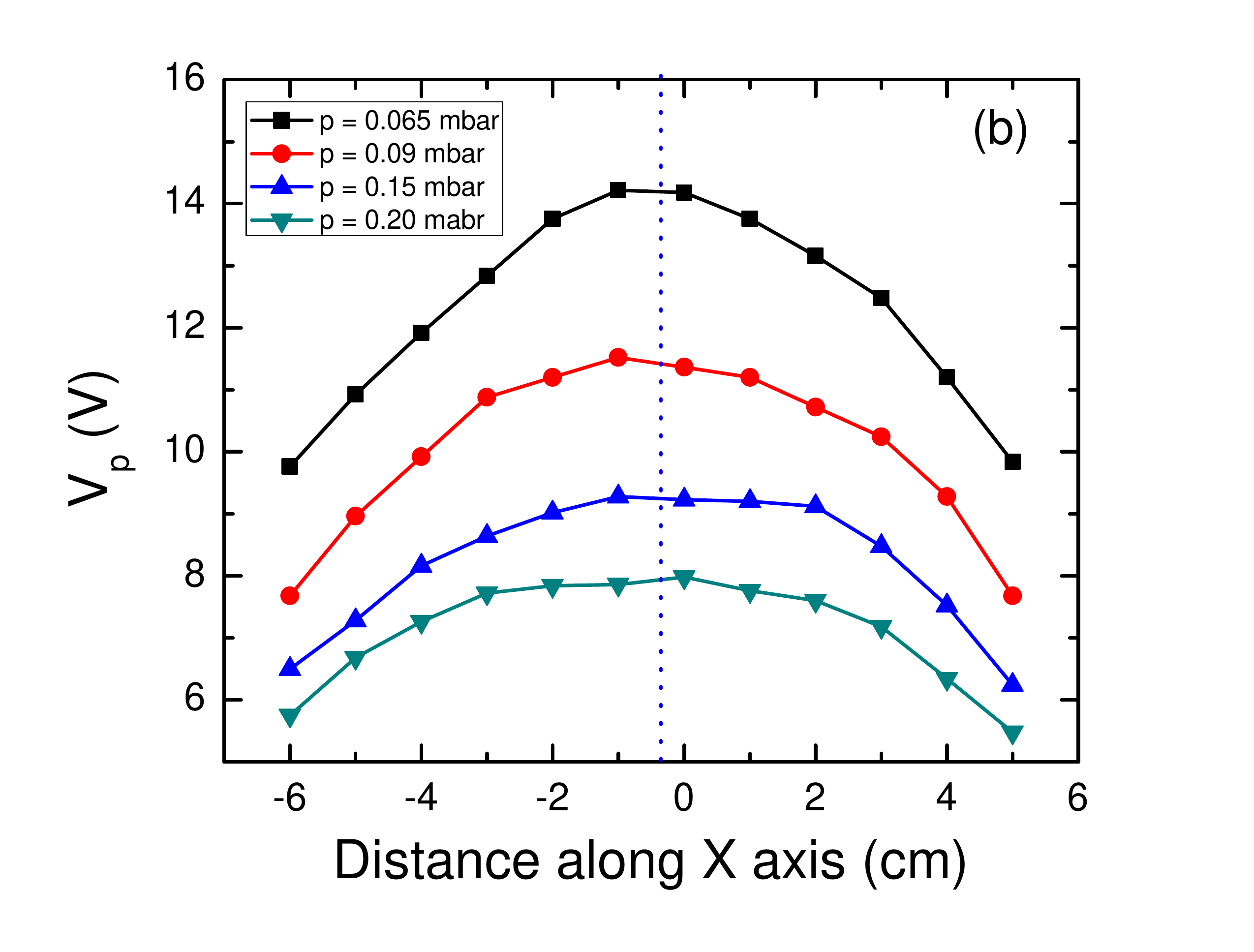}}}%
 \caption{\label{fig:fig8}(a) Plasma potential profiles along the X--axis at Y $\sim$ -3 $cm$ and Z $\sim$ 30 $cm$ for different rf powers at fixed pressure, p = 0.09 mbar (b) for different pressure at fixed rf power, P = 2.5 W. The errors in the measured value of plasma potential are within $\pm$ 2 V. All the measurements are taken in the absence of the dust particles.} 
 \end{figure*}
\par
 Fig.~\ref{fig:fig8}(a) shows the plasma potential ($V_p$) variation along the X-axis at Z $\sim$ 30 cm and Y $\sim$ -3 cm for different powers at constant pressure, p = 0.09 mbar. It is clear from the figure that dust cloud is symmetric about the Y axis at X $\sim$ 0 cm. The plasma potential varies monotonically on the both sides of the central region (X $\sim$ 0 cm), which gives rise to an E-field component in the X-direction. The plateau region of $V_p$  depends on the diffused length of plasma or input power, which reduces with decreasing the input rf power at fixed gas pressure. At higher power (P $>$ 4 W), edge particles doesn't exhibit the vortex motion. It is mainly due to small scale length of E-field in the dust cloud (or less extent of the dust cloud in the direction of $\beta$). It is predicted by Vaulina \textit{et al.} \cite{vaulinajetp,selfexcitedmotioninhomogeneus,vaulinasripta2004} that vortex motion is only possible when the width of the dust cloud crosses a threshold value or for a large number of layers. At below the threshold power, the E-filed diffuses to more dust layers; hence, dust grains start to rotate in the X--Y plane. It is also pointed out that vortex motion is independent of the charge gradient in the Y-direction \cite{vaulinaselfoscillation}. As shown in Fig.~\ref{fig:fig8}(a), the X--component of E--field is negligible (flat $V_p$) inside the central region of dust cloud (from X $\sim$ -1.5 to 1.5 $cm$); therefore, dust grains show random motion instead of the rotational motion. The effect of the gas pressure on the plasma potential along the X-axis is depicted in Fig.~\ref{fig:fig8}(b). It is clear that plasma potentials have a gradient on the both sides of the central axis (X $\sim$ 0 cm). The magnitude of the E-field decreases with increasing the neutral pressure. The plateau region of plasma potential increases with increasing the gas pressure, which is consistent with the larger dust cloud width  at higher power (see Fig.~\ref{fig:fig4}). At higher pressure (p $>$ 0.3 mbar), the E-field diffuses to a large number of layers; therefore, edge particles participate in the rotational motion. For the detailed descriptions of the vortices and its multiplicity, quantitative analysis to compare the experimentally observed results to an available theoretical model \cite{vaulinajetp,selfexcitedmotioninhomogeneus} are is provided for the set of discharge parameters.  
 \begin{figure*}
 \centering
\subfloat{{\includegraphics[scale=0.325]{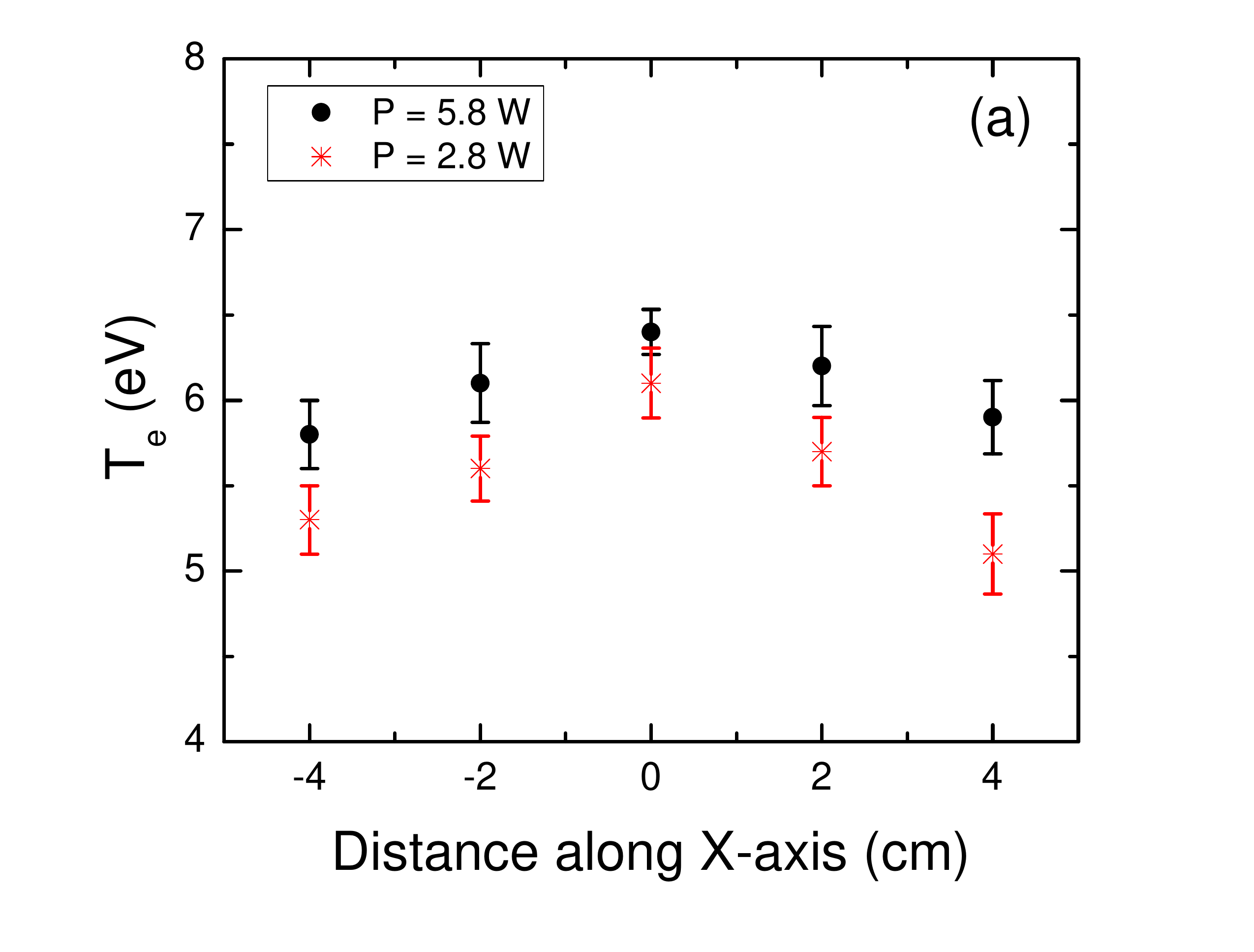}}}%
\hspace*{-0.6in}
 \qquad
\subfloat{{\includegraphics[scale=0.325]{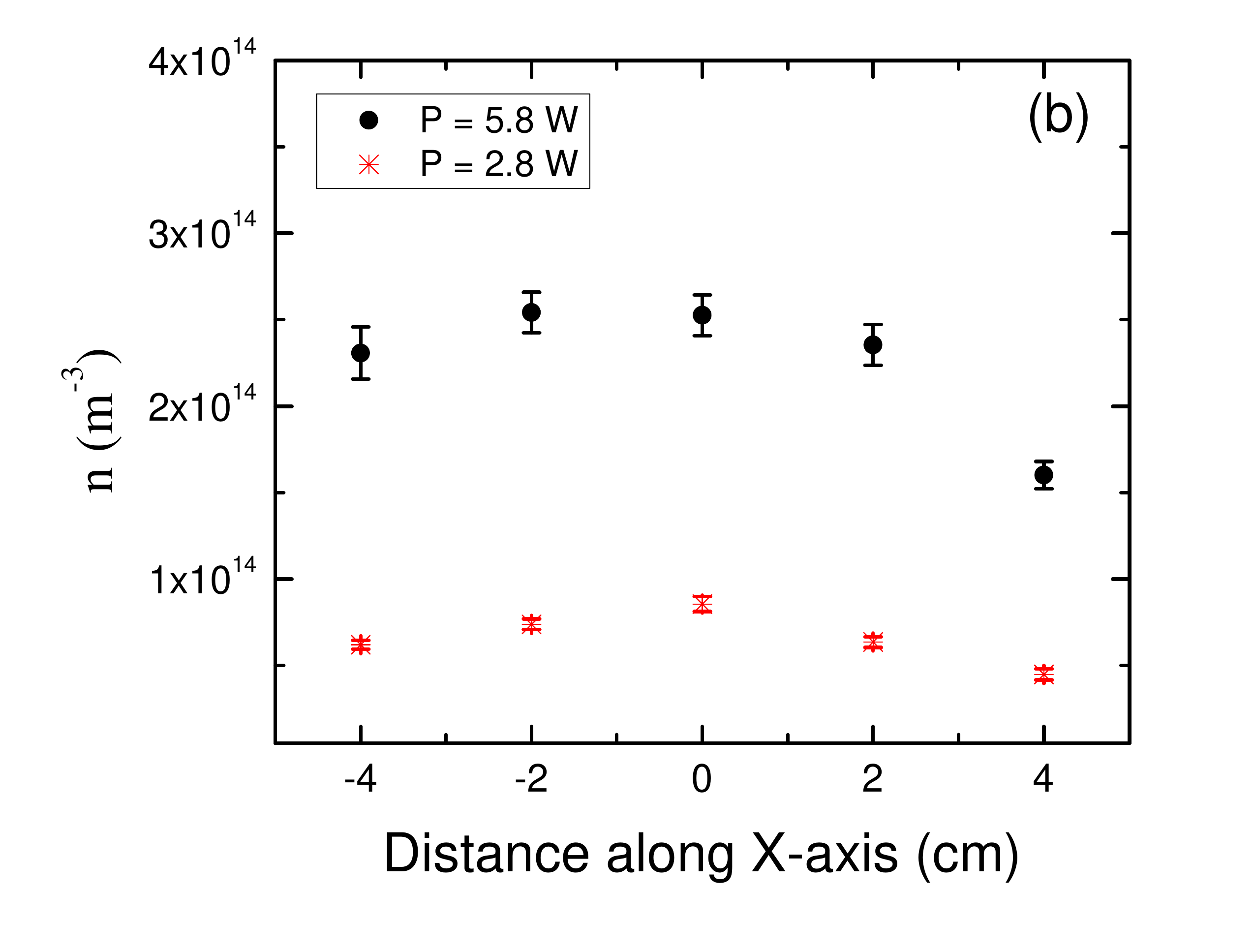}}}%
 \caption{\label{fig:fig9}(a) Electron temperature ($T_e$) variation along the X--axis at Y $\sim$ -3 $cm$ and Z $\sim$ 30 $cm$ for two rf powers P = 5.8 W and 2.8 W. (b) Plasma density ($n$) variation along the X--axis at Y $\sim$ -3 $cm$ and Z $\sim$ 30 $cm$ for two rf powers, P = 5.8 W and 2.8 W. All the measurements are taken in the absence of the dust particles at gas pressure, p = 0.12 $mbar$.}
 \end{figure*}  
 \par
Fig.~\ref{fig:fig9}(a) shows the electron temperature variation along the X-axis at Y $\sim$ -3 cm and Z $\sim$ 30 cm for input powers, P = 5.8 W and 2.8 W at gas pressure, p = 0.12 mbar. The $T_e$ is observed to be high at the axis passes through the centre (at X = 0 cm) and decreases on the both sides of X = 0 line. It essentially indicates the inhomogeneity (or gradient) in $T_e$ form the central axis. The values of gradient in $T_e$ depend on the discharge parameters. The similar trend of $T_e$ is also observed for the pressure regime (p = 0.06 to 0.3 mbar) and power regime (P = 4 W to 2.5 W). Similar to $T_e$, plasma density ($n$) is measured along the X-axis at Y $\sim$ -3 cm and Z $\sim$ 30 cm for same discharge parameters. Typical plasma density variation along the X-axis for two powers at fixed pressure is shown in Fig.~\ref{fig:fig9} (b). Similar to the $T_e$, plasma density varies monotonically on the both sides of central axis. The finite gradient in $T_e$ and $n$ on the both sides of central axis (X = 0 cm) is mainly responsible for the charge gradient on the both sides of dust cloud. The charge gradient is estimated using the expression of Eq.(2), which strongly depends on the gradient of $T_e$. As is known that dust charging mechanism depends on the dust Debye length. The dust grains have higher negative charge in high and lower in low plasma density plasma region. However, the effect of density gradient on the dust charge gradient is negligible for the given discharge conditions.
\par
To compare the experimentally observed angular frequency ($\omega_{exp}$) with the theoretically predicated value ($\omega_{th}$), the charge gradient ($\beta$) along the X-axis (from X =0 cm) is estimated for the given discharge parameters (P = 2.8 W and p = 0.12 mbar). The charge gradient is estimated by using the relation, $\nabla Q_d = (Q_{X2} - Q_{X1})/(X_2-X_1)$,
 where $X_1$ and $X_2$ are the two spatial points on the dust cloud. For the quantitative analysis, only average sized particles ($\sim$ 2 $\mu$m) are considered based on the force balance conditions. To measure the $\omega_{exp}$, we have used vortex-I and Vortex-II (as indicated in Fig.~\ref{fig:fig4}(a)) of the dust cloud. The observed value of $\omega_{exp}$ is varied between 0.4 to 0.9 rad/sec for the same discharge parameters. Theoretically estimated value of angular frequency ($\omega_{th}$) comes out to be $\sim$ 0.7 to 1 rad/sec for $\beta/e Z_0 \sim  $ 0.03 to 0.04 $cm^{-1}$, $M_d \sim$ 8 $\times 10^{-14}$ kg, $g$ = 980 $cm^{-2}/sec$ and $\nu_{dn} \sim$ 24 $sec^{-1}$. 
 It can be concluded that the experimentally measured values of angular frequency and the theoretically predicted values (by Vaulina \textit{et al.}\cite{vaulinajetp}) are in good agreement for the given discharge parameters. Moreover, the direction of rotation of particles is also consistent with the direction predicted by the theoretical model \cite{selfexcitedmotioninhomogeneus,vaulinajetp}.
 \par
 The characteristic size $D_0$ of a vortex structure can be obtained from the viscosity ($\eta_k $) of the dusty plasma medium \cite{vaulinasripta2004}, which can be  expressed as $D_0 = \alpha \left(\eta_k / (\omega^* + \nu_{dn})\right)^{1/2}$, where $\omega^*$ is effective dusty plasma frequency and $\alpha$ takes into the difference between viscosity in quasi--stationary and dynamic vortex structure. The coefficient $\alpha$ is estimated as $\approx$ 49 \cite{vaulinasripta2004}. The variation of kinetic viscosity ($\eta_k$) with a wide range of discharge parameters and coupling constant ($\Gamma$) is discussed by Fortov \textit{et al.}\cite{fortovviscosity2}. For the present set of experiments, the dusty plasma medium is assumed to be in liquid state; therefore, an effective coupling constant ($\Gamma^*$) \cite{vaulinasripta2004} has the value $<$ 170. For this range of coupling parameters, the kinetic viscosity $\eta_k$ is considered to be $\sim$ 0.02 to 0.04 $cm^2 s^{-1}$ similar to the value reported in Refs.\cite{fortovviscosity2,nosenkoviscosity2}. For the discharge parameters (P = 2.8 W, p = 0.12 mbar), the estimated characteristic size ($D_{th}$) of the vortex structures for $\eta_k$ = 0.02--0.04 $cm^2 s^{-1}$ and $\nu_{dn}\sim$ 24 $s^{-1}$ comes out to be $\sim$ 3--5 $mm$, which is in good agreement with the experimentally measured vortex diameter, $D_{exp} \sim$ 4-5 $mm$ (see Fig.~\ref{fig:fig4}(a)). Similarly, the vortex size is measured for other discharge conditions (for p = 0.2 mbar to 0.3 mbar), which matches well with the theoretical estimation. Since, the dimension of dust cloud is multiple of the dust vortex size; therefore, a series of the vortex structures is formed on the both sides of the dust cloud in the X--Y plane. The number of vortex structures strongly depends on the aspect ratio of the dusty plasma medium. For a constant width of dust cloud, multiple vortex structures are observed with increasing the length of the dust cloud, which essentially signifies that large aspect ratio dusty plasma can accommodate the multiple vortices. 
 \par
\begin{figure}
\centering
  \includegraphics[scale= 0.300]{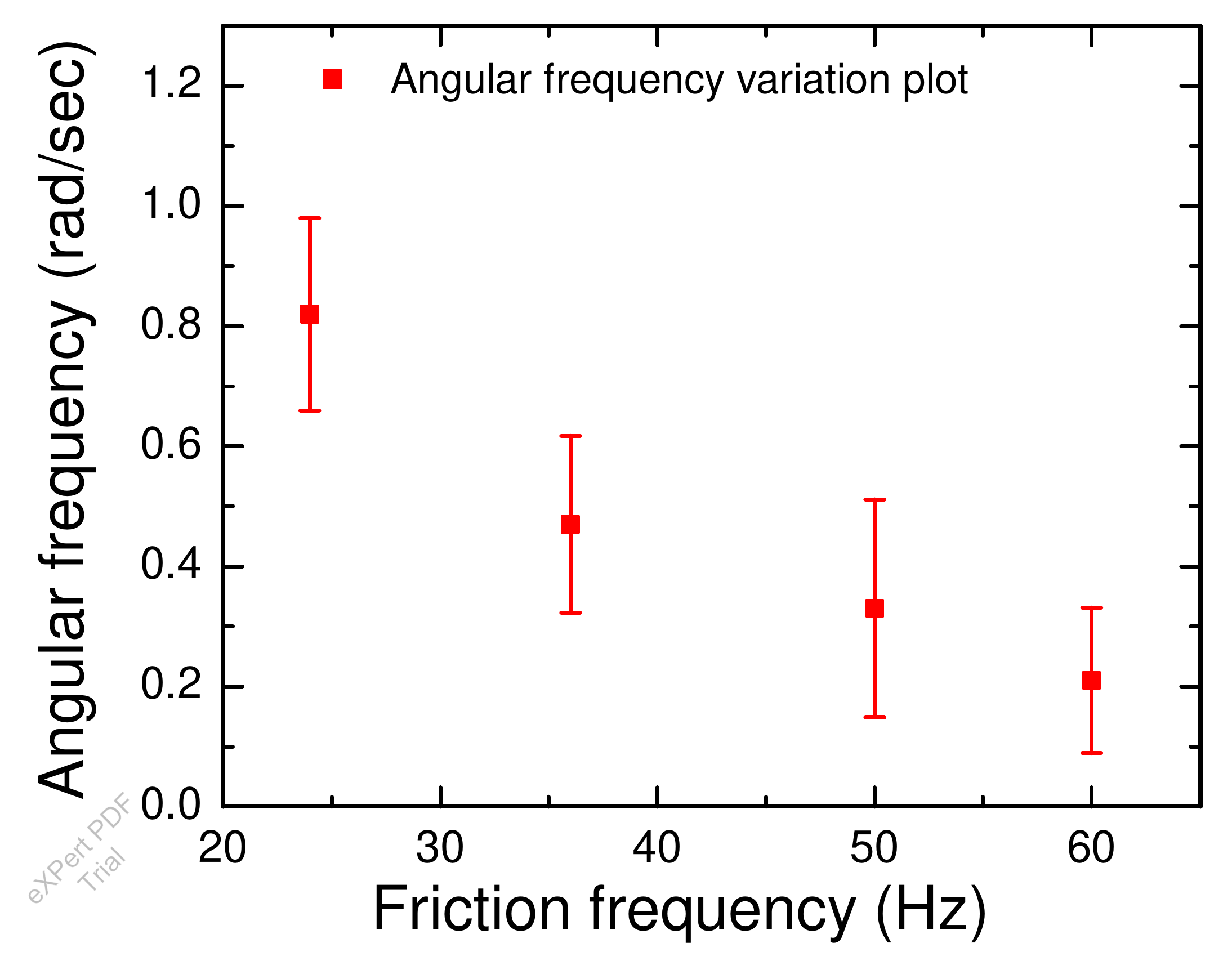}
\caption{\label{fig:fig10} Angular frequency ($\omega_{exp}$) variation of dust particles with friction frequency at P = 2.8 W}  
 \end{figure}

The dependence of friction frequency ($\nu_{dn}$) on the angular velocity of the rotating dust grains is depicted in Fig.~\ref{fig:fig10}. An average angular frequency ($\omega$) of the dust grains decreases  with increasing the dust-neutral friction. It essentially specifies that dissipative instability, which gives rise to vortex motion, is independent of the dissipation losses of the dust grain medium. These observed results are similar to the theoretical prediction of Vaulina \textit{et al.} \cite{vaulinajetp,selfexcitedmotioninhomogeneus,vaulinaselfoscillation} model. It concludes that the charge gradient of dust grains along with the gravity is a mainly possible source to excite the vortex motion. 
 
\section{Summary and conclusion}  \label{sec:conclusion}
The dynamics of dust grains in a large aspect ratio dust plasma in the background of inhomogeneous plasma over a wide range of discharge parameters are studied. Inductively coupled discharge is initiated in the background of argon gas in the source tube, which later diffuses in the main experimental chamber. The diffused plasma is characterized using the electrostatic probes to understand the dynamics of dust grains. For injecting the dust grains into the potential trap, a novel technique using the DC glow discharge is employed. In the DC plasma background, the dust grains acquire negative charges on their surface and levitate at the sheath-plasma interface. Due to the poor confinement, these charged grains leave the confined region and come into diffused plasma volume. These grains then start to flow under the action of an axial ambipolar electric field of the diffused plasma and confined in the electrostatic trap. In this trap, particles are confined under the combined action of electrostatic forces, which are due to the diffused plasmas (ambipolar E--field) and plate charges (sheath E--field), and gravitation force. The dynamics of dust grain medium is recorded in the X-Y plane at Z $\sim$ 30 cm for various discharge conditions. The main findings of the experimental studies are listed below,
\begin{enumerate}
 \item  The inductively coupled diffused plasma is used to create a large volume or large aspect ratio dusty plasma at low pressure.
 \item The dusty plasma medium exhibits waves like motion at low pressure and it is strongly dependent on the friction frequency of particles with neutrals.
 \item The clockwise and anti-clockwise series of co-rotating vortices are observed on the both sides of dust cloud at lower power and higher pressure. The vortex motion is independent of the friction frequency or dissipation losses of the medium.
\item The co-rotating vortex series are only observed above a threshold dust cloud width. 
\item Multiplicity of vortex depends on the dimension (or length) of the dust cloud in a given 2D plane.
\end{enumerate}
 The angular frequency of the rotation based on the model provided by Vaulina \textit{et al.} \cite{vaulinajetp,selfexcitedmotion} is found to be in close agreement with the experimentally observed values, which essentially indicates that the charge gradient of dust particles orthogonal to the gravity is a possible mechanism to drive the vortex flow. Vaulina \textit{et al.}\cite{vaulinajetp,selfexcitedmotioninhomogeneus} have also pointed out that a small charge gradient in the dust cloud ($\sim$ 1\%) is an effective source for conversion of potential energy to kinetic energy of dust grains. The occurrence of charge gradient is due to the plasma inhomogeneity from the central region to the outer edges of the dust cloud. In the dissipative medium, vortex has the characteristic size; therefore, a series of the co-rotating vortex on each side of dust cloud is observed. The multiplicity of the vortex strongly depends on the dimension of the dust cloud. The present studies focus on the dynamical studies of large aspect ratio (2D) dusty plasma. However, the detailed collective dynamics of the large volume (3D) dusty plasma is still under investigation and will be reported in the future publications.
 \section{Acknowledgement} 
The authors grateful to Dr. M. Bandyopadhyay for his suggestions and invaluable inputs to improve the manuscript.
\bibliography{aipsamp}
\end{document}